\documentclass[a4paper,conference]{IEEEtran}
\IEEEoverridecommandlockouts

\usepackage{cite}

\usepackage{amsmath,amssymb,amsfonts}
\usepackage{graphicx}
\usepackage{booktabs}
\usepackage{url}
\usepackage{algorithmic}
\usepackage{subcaption}
\begin{document}

\title{Satellite-Based Quantification of Contrail Radiative Forcing over Europe: A Two-Week Analysis of Aviation-Induced Climate Effects}

\author{
\IEEEauthorblockN{
    Irene Ortiz, Javier García-Heras, Manuel Soler,\\ Amin Jafarimoghaddam
}
\IEEEauthorblockA{
    University Carlos III of Madrid \\
    Madrid, Spain \\
    \small irortiza@ing.ucm3.es, gcarrete@ing.uc3m.es,\\ masolera@ing.uc3m.es,  ajafarim@pa.uc3m.es
}
\and
\IEEEauthorblockN{
    Ermioni Dimitropoulou,Pierre de Buyl, Nicolas Clerbaux
}
\IEEEauthorblockA{
    Royal Meteorological Institute of Belgium \\
    Brussels, Belgium \\
    \small ermioni.dimitropoulou@meteo.be,pdebuyl@meteo.be, \\  nicolas.clerbaux@meteo.be
}
\and
\IEEEauthorblockN{
    Hugues Brenot, Jeroen van Gent
}
\IEEEauthorblockA{
    Royal Belgian Institute for Space Aeronomy \\
    Brussels, Belgium \\
    \small brenot@aeronomie.be,  jeroen.vangent@aeronomie.be
}
\and
\IEEEauthorblockN{
    Klaus Sievers
}
\IEEEauthorblockA{
    Klaus Sievers Services \\
    Munich, Germany \\
    \small Sievers.AvWx@Web.de
}
\and
\IEEEauthorblockN{
    Evelyn Otero, Parthiban Loganathan
}
\IEEEauthorblockA{
    KTH Royal Institute of Technology \\
    Stockholm, Sweden \\
    \small otero@kth.se, loganathan.parthiban@gmail.com
}
}

\maketitle

\begin{abstract}
Aviation's non-CO$_2$ effects, especially the impact of aviation-induced contrails, drive atmospheric changes and can influence climate dynamics. Although contrails are believed to contribute to global warming through their net warming effect, uncertainties persist due to the challenges in accurately measuring their radiative impacts. This study aims to address this knowledge gap by investigating the relationship between aviation-induced contrails, as observed in Meteosat Second Generation (MSG) satellite imagery, and their impact on radiative forcing (RF) over a two-week study. Results show that while daytime contrails generally have a cooling effect, the higher number of nighttime contrails results in a net warming effect over the entire day. Net RF values for detected contrails range approximately from -8 TW to 2.5 TW during the day and from 0 to 6 TW at night. Our findings also show a 41.03 $\%$ increase in contrail coverage from January 24–30, 2023, to the same week in 2024, accompanied by a 128.7 $\%$ rise in contrail radiative forcing (CRF), indicating greater warming from the added contrails. These findings highlight the necessity of considering temporal factors, such as the timing and duration of contrail formation, when assessing their overall warming impact. They also indicate a potential increase in contrail-induced warming from 2023 to 2024, attributable to the rise in contrail coverage. Further investigation into these trends is crucial for the development of effective mitigation strategies.
\end{abstract}

\begin{IEEEkeywords}
Contrail Radiative Forcing, Non-CO$_2$ effects, Global Warming, Meteosat Second Generation, Satellite Data
\end{IEEEkeywords}

\section{Introduction}
adiative forcing (RF) measures changes in Earth's energy balance resulting from various factors, including greenhouse gases, aerosols and cloud formations, and is essential for understanding climate dynamics. Contrails—high-altitude clouds composed of ice particles and generated by aircraft emissions—affect this balance through two primary mechanisms: shortwave and longwave radiative effects. The shortwave effect occurs when contrails reflect incoming solar radiation back into space, resulting in a cooling effect on the Earth's surface. In contrast, the longwave effect arises as contrails trap outgoing infrared radiation, which would otherwise escape into space, thereby contributing to a warming effect. The net Contrail Radiative Forcing (CRF), which is the main metric for assesing contrail's climate impact, is obtained by systematically aggregating these opposing effects in each contrail feature.  \\\\
While numerous studies have consistently shown that contrails contribute to a net warming effect on a global, annual scale \cite{lee_contribution_2021, chen_radiative_2000, burkhardt_mitigating_2018}, the radiative impact of individual contrails remains less well understood. To better quantify these effects, some studies have focused on single-case analyses to estimate CRF over the full lifespan of individual contrails. For example, \cite{haywood_case_2009} reported a contrail persisting for 18 hours near the UK with a radiative effect ranging from 30 W/m² at night to 10 W/m$_2$ during daylight hours. More recently, Xingue Wang et al. \cite{wang_radiative_2024} examined a two-day contrail-cirrus outbreak over Western Europe (June 22–23, 2020), estimating an accumulated CRF of 2.3 TW over a 24-hour period. Other studies have taken a broader approach by estimating the RF from aviation-induced contrails across specific regions to provide a more comprehensive understanding of the overall impact. For instance, in \cite{schumann_aviation-induced_2013}, the authors assessed the longwave and shortwave CRF using data from the Meteosat-9 satellite's Spinning Enhanced Visible and InfraRed Imager (SEVIRI), which provided extensive coverage over a 10-hour period. They proposed distinguishing aviation-induced cirrus by correlating it with air traffic patterns and estimated a net CRF of approximately 40-80 mW/m$_2$. These studies have improved our understanding of the radiative effects of contrails under various conditions. However, their limited scope—concentrating on specific or short-term events—makes it difficult to draw general conclusions. A thorough understanding of contrails' contribution to aviation-induced climate change necessitates quantifying CRF over larger geographic areas and extended time periods. \\\\In this study, we assess the radiative impact of contrails over an extensive geographic area and a prolonged continuous period using satellite measurements and weather data. Specifically, we analyzed two full weeks—January 24–30, 2023, and January 24–30, 2024—focusing on all contrails within the complete field of view of the Meteosat Second Generation (MSG) satellites. To accomplish this, we employed a Contrail Detection model \footnote{For further details, please refer to our forthcoming paper on Contrail Detection using geostationary satellite imagery.}
 trained to identify all visible contrails in some thermal infrarred satellite bands. Shortwave and longwave RF values for each scene are determined through multidimensional interpolation using pre-existing Look-Up Tables (LUTs) \footnote{For further details, please refer to our forthcoming paper on Radiative Forcing Estimations and LUTs.}. These LUTs encompass radiative data for a wide range of cloud parameters and various conditions, including different solar zenith angles, land cover types, and sea and surface temperatures. The automation of contrail detection has allowed us to analyze over 700,000 contrails (counting as distinct instances those observed at different time steps) across 1,344 scenes covering Europe, North Africa, and the surrounding oceans. This analysis broadens our understanding of contrail warming impacts and sets the stage for a detailed year-long study to more precisely quantify contrail radiative forcing effects.
\\\\The document is structured as follows: Section 2 details the sources of data used to produce the experiment; Section 3 describes the methods employed detect contrails in satellite images and find their radiative properties; Section 4 presents the analysis of the results obtained; and Section 5 offers conclusions and outlines directions for future research.

\section{Data}
The experiment was conducted over two complete weeks, encompassing 24-hour periods each day. The selected periods were from January 24th to January 30th in both 2023 and 2024. Notably, exceptionally warm temperatures were reported in January 2024. While multiple factors could contribute to the observed temperature increase, visual inspection in the satellite imagery revealed a significant rise in both the number of contrails observed and those detected by models during this period. This increase in contrail coverage could be a potential contributor to the warming reported.\\\\The data used in this work was obtained from the SEVIRI onboard the MSG satellites of the European Organisation for the Exploitation of Meteorological Satellites (EUMETSAT). In particular, the data from the MSG-3 (Meteosat-10) and MSG-4 (Meteosat-11) satellites have been used for this work. Positioned in geostationary orbit, about 36,000 kilometres above Earth, these satellites provide spectral information across 11 channels, including visible and infrared regions \cite{schmetz_introduction_2002}.  Observations are captured every 15 minutes with a spatial resolution of approximately 3x3 km² at the subsatellite point, with a pixel size that grows as one looks further from the equator. The SEVIRI level 1b data serves two main purposes: (a) generating the false-color RGB images for contrail detection models, and (b) providing input to the Optical Cloud Analysis (OCA) system \cite{king_cloud_nodate}, which we use for the physical characterization of the clouds. These physical parameters serve as the input to the radiative forcing estimation. The specifics of this retrieved information are as follows:

\begin{enumerate}
    \item \textit{Ash Composite}: This product generates false-color RGB images to enhance contrail visibility by combining several MSG thermal infrared (IR) bands. It is composed of the red Brightness Temperature (BT) difference  IR12µm-IR10.8µm to highlight contrails by their higher transmissibility compared to natural cirrus, the green BT difference IR10.8µm-IR8.7µm to differentiate cloud phases, and the blue BT IR10.8µm to accentuate contrails by leveraging their colder temperatures relative to surrounding features.
    See \cite{noauthor_eumetsat_nodate} for the definition. The Ash RGB composite was also used in \cite{ng_contrail_2024} for detecting contrails.\\
    \item \textit{Cloud parameters:} Cloud state parameters are characterized by cloud phase (CP), cloud top pressure (CTP), cloud optical thickness (COT), and cloud effective radius (CER). These parameters are obtained from the OCA of EUMETSAT, which employs the Optimal Estimation (OE) method along with SEVIRI spectral measurements simultaneously.  The cloud information obtained can be separated into upper layer and lower layer clouds. The upper layer consists of ice clouds, and includes values both for cases where the ice clouds are alone and where they coexist with underlying water clouds. The lower layer shows  data only when a water cloud is present beneath the upper ice layer.\\
    \item \textit{Forecast Data}: We use the skin temperature from Numerical Weather Prediction (NWP) data obtained from the European Centre for Medium-Range Weather Forecasts (ECMWF) \cite{noauthor_ecmwf_nodate}.\\
    \item \textit{Land Cover Data}: We use the MODIS L3 500m Land Cover dataset MCD12Q1 v061 \cite{friedl_mcd12q1_2019}.
\end{enumerate}

We resample all input to a regularly spaced grid in latitude and longitude with a grid spacing of 0.04 degrees.

\section{Methods}
The methodology employed to quantify the RF of contrail cirrus detectable by a geostationary satellite involves three key steps: First, a contrail detection model identifies the locations of all visible contrails within the scene. Second, the RF of all clouds present in the image is calculated. Third, the results from these two steps are intersected to determine the RF specifically attributable to the detected contrails.

\subsection{Radiative Forcing Calculations}
\begin{figure}[h]
    \centering
    \includegraphics[scale=0.39]{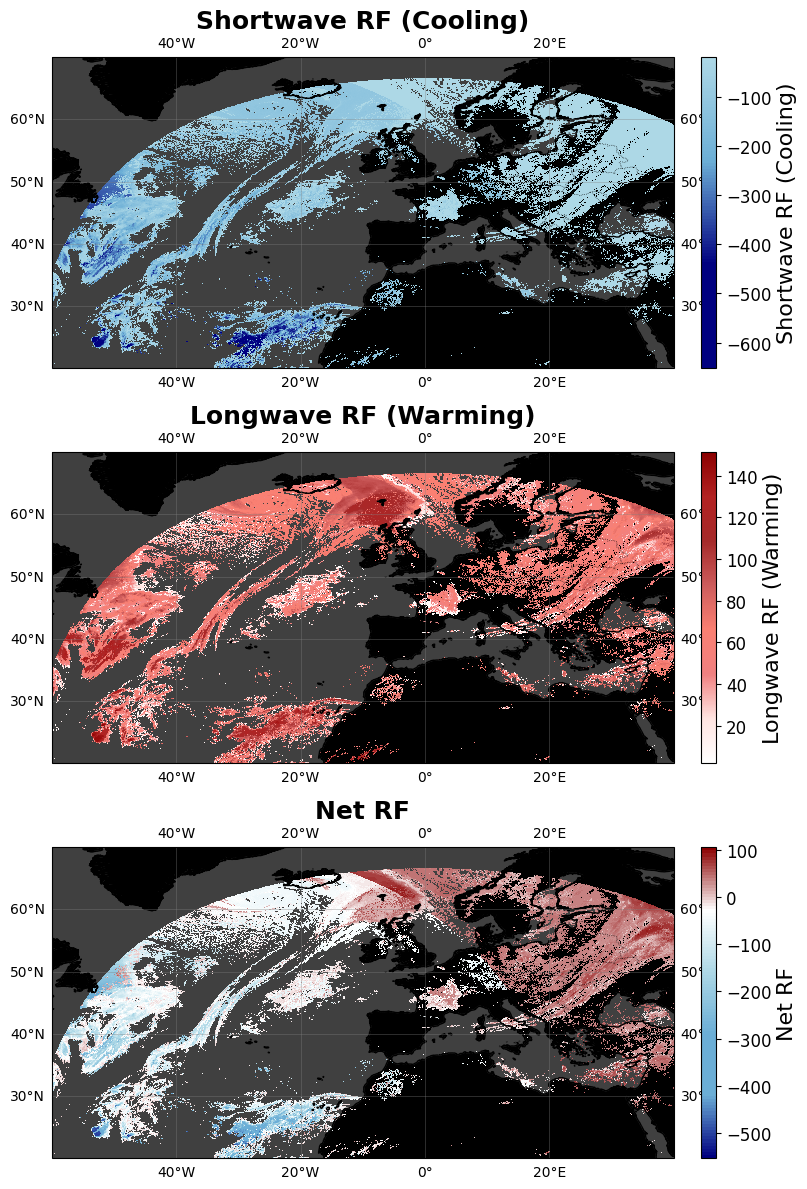}
    \caption{Radiative Forcing (RF), measured in Watts per square meter, in the Short Wave (SW) (first image) and in the Long Wave (LW) (second image) for the field of view of Meteosat satellites on January 24, 2023, at 08:00 UTC. The final image illustrates the net RF obtained by aggregating the SW and the LW RF.}
    \label{rf_estimation}
\end{figure}

The radiative forcing estimation was derived using multi-dimensional interpolation on pre-built Lookup Tables (LUTs). These LUTs were constructed using the libRadtran  radiative transfer library\cite{emde_libradtran_2016} to simulate both shortwave and longwave radiative forcing for various combinations of thin to semi-transparent ice cloud parameters, along with other relevant factors like solar zenith angles, sea or land surface temperature, surface type and the presence of an underlying water cloud. By interpolating the simulated shortwave and longwave RFs from the LUTs as a function of the surface type, geometry, skin temperature, and OCA parameters for each pixel, a detailed and location-specific assessment of radiative forcing is provided. This approach streamlines the radiative forcing estimation process, removing the need to repeatedly run time-consuming radiative transfer simulations in future large-scale analyses. Figure \ref{rf_estimation} provides an example of the shortwave, longwave, and net radiative forcing estimates obtained through this process. We use the sign convention of downward flux, so that a positive value of the RF represents a warming effect.
\subsection{Contrail Detection}
For this experiment, we utilized a single-frame U-Net-based network previously trained on the OpenContrails Dataset \cite{ng_opencontrails_2023}, which comprises approximately 22,000 images captured by the Geostationary Operational Environmental Satellites - 16 Series (GOES-16) \cite{noauthor_images_nodate} between April 2019 and April 2020 in different locations of North and South America. This dataset has been selected for training due to its substantial number of labelled scenes (55\%), necessary for capturing all the variability of contrail features.\\
\subsubsection{Network Architectural Details}
The architecture employed is a hybrid neural network that combines a transformer-based encoder with a convolutional decoder. Because of the limitations in the computational resources, the encoder is a lightweight variant of the CoaT (Co-Scale Conv-Attentional Image Transformers) \cite{xu_co-scale_2021} model, called CoaT-Lite Mini. This variant is optimized for efficient image processing, avoiding parallel blocks and incorporating a reduced channel depth in each layer. The model processes images through four sequential blocks, where feature maps are downsampled and converted into image tokens. These tokens are analyzed using convolutional operations for local pattern extraction and self-attention for capturing image inter-part relationships. The output from each block is reshaped into a 2D feature map and forwarded to the next block and decoder via skip connections. The decoder employs sequential convolutional blocks with upsampling, producing feature maps at three different resolutions. The three feature maps are combined using a Feature Pyramid Network (FPN) \cite{lin_feature_2017}. The output is then regularized with a dropout layer, applying a 0.5 probability to deactivate weights to prevent overfitting to the training data. Finally, the features are upsampled to the original image size, with the number of channels reduced to one. Each pixel in this single-channel output mask represents the probability of being part of a contrail. The optimization of the weights of the network was performed using the AdamW optimizer \cite{loshchilov_decoupled_2019} over 30 epochs, minimizing a convex surrogate of the Dice loss function. Transfer learning was used to initialize the encoder's weights with those from a CoaT network pretrained on ImageNet \cite{deng_imagenet_2009}.
\begin{figure*}[!t]
    \centering
    \includegraphics[scale=0.32]{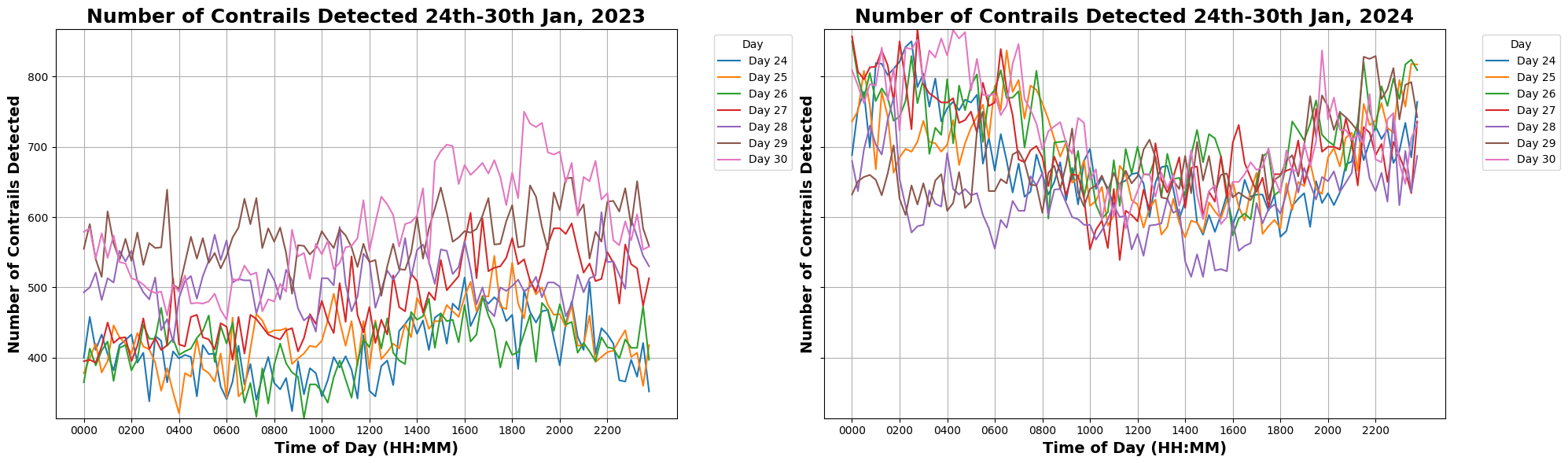}
    \caption{Comparison of the number of contrails detected across the entire field of view of Meteosat satellites in the week of the 24th-30th of January of 2023, and the same week in 2024, encompassing all times of day.}
    \label{conts}
\end{figure*}
\subsubsection{Domain Adaptation}
The trained network was used to detect contrails in MSG Ash RGB images, which have different characteristics from the original training data, especially in terms of geographical coverage and image resolution. MSG images cover Europe, Africa, and portions of the Atlantic Ocean, with a maximum thermal infrared resolution of 3x3 km². In contrast, the Ash RGB training images from GOES-16 focus on the United States, the Atlantic Ocean, South America, and the Caribbean, offering a finer resolution of 2x2 km² at nadir. Although the difference in geographical coverage is not expected to have an impact, the disparity in resolution could affect detection accuracy. To address this, the resolution of MSG images was adjusted using bilinear interpolation, simulating a 2x2 km² resolution at nadir. Given the large size of an MSG scene, a sliding window of 256x256 pixels was applied to divide the images into smaller overlapping sections. The detector was applied to each section, with results from overlapping areas combined to ensure accurate identification of objects partially visible across sections. Finally, after aggregating all the detections produced by the sliding window, the contrail mask for the entire scene was transformed back to the original resolution, preserving the true sizes of the segmented contrails.

\section{Results}

This section presents our results, starting with cumulative RF and CRF calculations every 15 minutes over two weeks to assess yearly variations. We then analyze correlations with cloud parameters and other factors. 
\subsection{Yearly Comparison}
The analysis focuses on examining the changes observed during the two selected weeks. To draw more definitive conclusions, we plan to extend the study to cover a longer time frame in a future experiment. This extension will help account for seasonal variations and other factors that may influence the results.\\
\subsubsection{Changes in Contrail Coverage}
We compare the features detected by the model for the same week in 2023 and 2024 to assess changes in contrail coverage. Figure \ref{conts} shows the number of contrails detected at various times throughout the week, revealing a significant increase in 2024. Specifically, there is a 41.03\% rise in total contrail detections compared to 2023. This increase aligns with heightened passenger traffic reported by the International Air Transport Association (IATA) in their January 2024 Air Passenger Market Analysis \cite{vanolli_air_2024}, which noted a 10\% increase in Revenue Passenger Kilometers (RPKs) and a 9.6\% increase in Available Seat Kilometers (ASKs) for European flights.\\
\subsubsection{Changes in Contrail Warming}
We now analyze the changes in RF to determine if the observed increase in contrail coverage is associated with a rise in total warming.\\
\begin{figure}[h]
    \centering
    \includegraphics[scale=0.35]{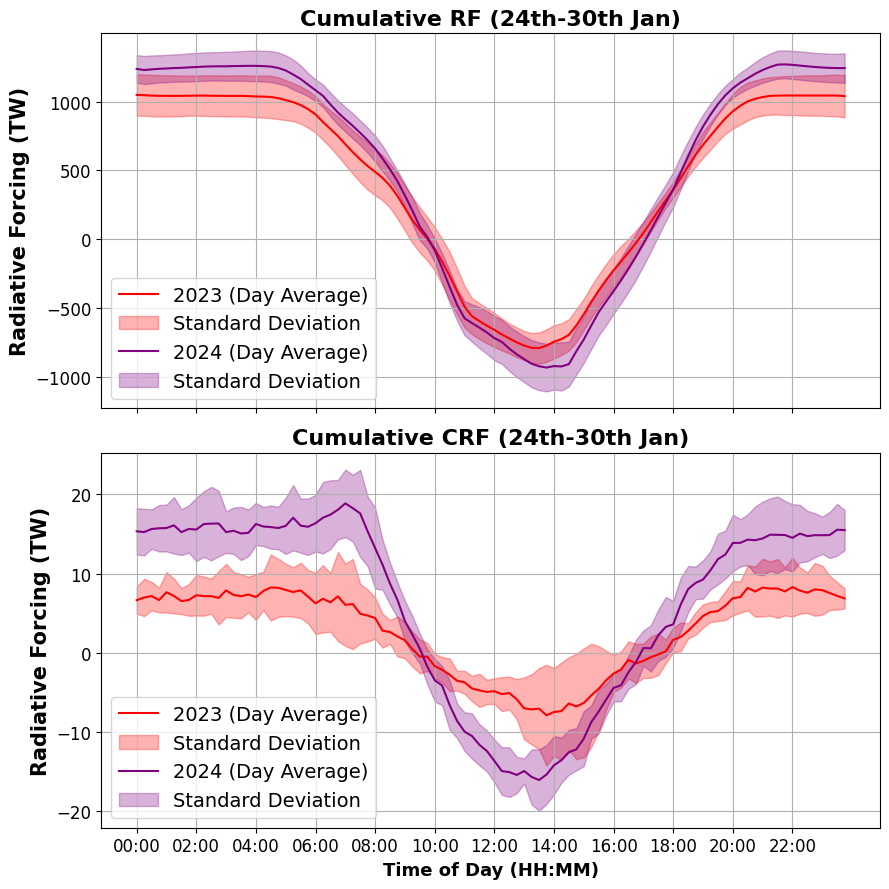}
    \caption{Comparison of the cumulative Radiative Forcing (RF) curves (top) and the cumulative contrail RF curves (bottom) between the week of the 24th-30th of January of 2023 and the corresponding week in 2024. }
    \label{dayly_evolution}
\end{figure}

The cumulative RF and CRF values shown in Figure \ref{dayly_evolution} were derived by aggregating the net RF of all clouds (top) and the net CRF of all contrails detected (bottom) across the entire field of view of the MSG satellites. The values for each time of day are averaged over the seven-day period.
\\\\The analysis indicates that both total cloud RF and CRF exhibited more extreme values in 2024 compared to 2023, with increased cooling (more negative RF/CRF values) during the day and heightened warming (more positive RF/CRF values) at night. Given the observed cooling effect during the day and the warming effect at night, we aggregate the total forcing across all times of day to estimate the overall warming effect during one complete day (see Table \ref{rf_comparison}). The data reveals two key findings: (1) the combined effect of nighttime warming and daytime cooling results in a net warming effect in both years, and (2) there is an increase in both cloud and contrail warming during the week of 2024. Specifically, total cloud RF rose by $\textbf{19.51\%}$, while total CRF surged by $\textbf{128.7\%}$, indicating that the increased warming in 2024 is affected by the rise in contrail coverage. However, further research in the aggregating factors for this increased warming is required.
\begin{table}[ht!]
  \begin{center}
    \caption{Average daily RF and CRF values for the 24-30th Jan, 2023 and the 24-30th Jan, 2024}
    \label{rf_comparison}
    \begin{tabular}{@{}lcr@{}}
      \toprule 
      \textbf{Year} & \textbf{Cumulative RF } & \textbf{Cumulative CRF } \\ 
      \midrule 
      2023 & $41,000 $ TW & $258$ TW \\
      2024 & $49,000$ TW & $590$ TW \\
      $\Delta(2024-2023)$  & $\textbf{19.51\%}$ & $\textbf{128.7\%}$ \\
      \bottomrule 
    \end{tabular}
  \end{center}
\end{table}
\subsection{Analysis of Individual Contrails}
\begin{figure}[h]
    \centering
    \includegraphics[scale=0.24]{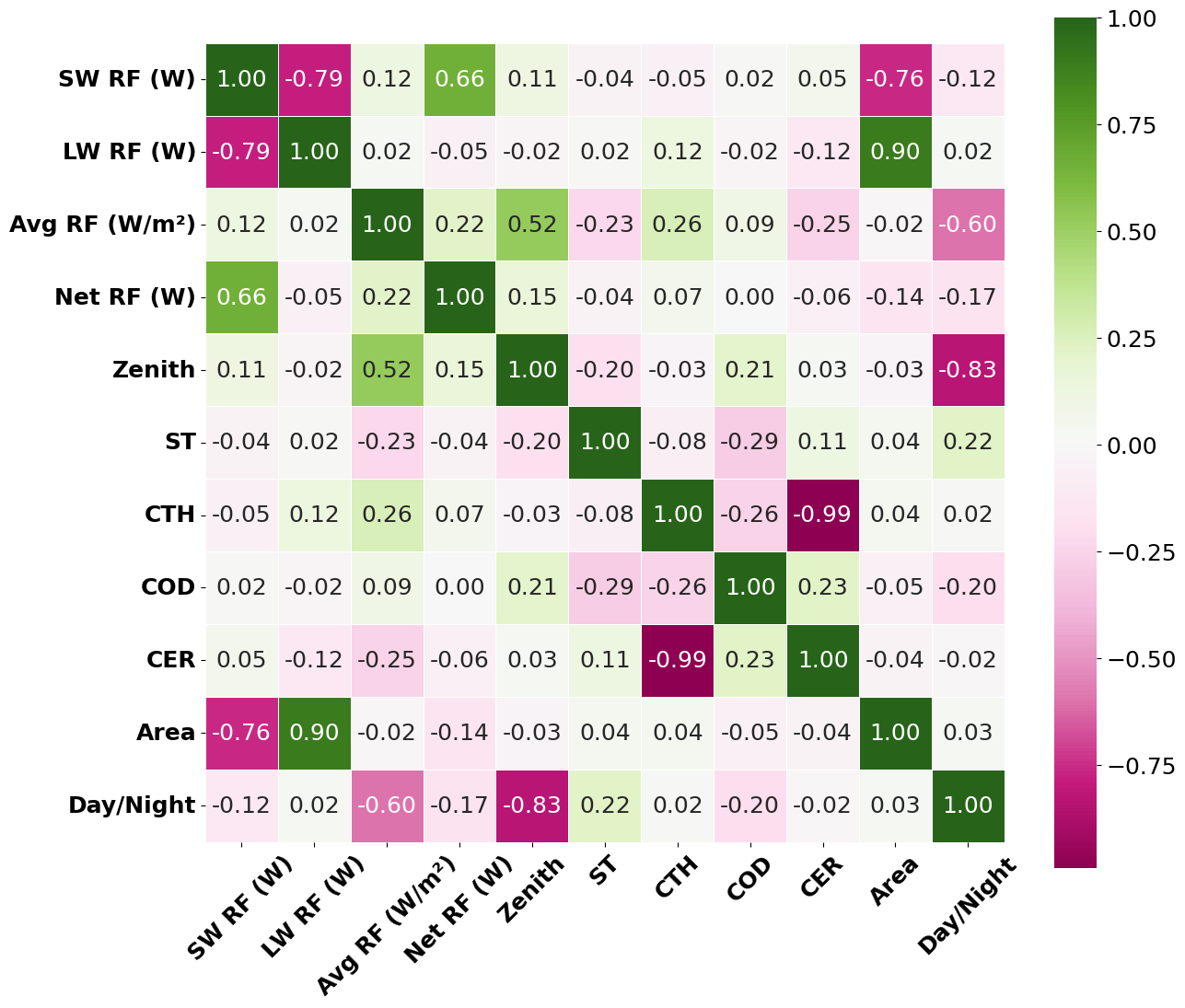}
    \caption{Correlation Matrix between contrail cloud parameters (COD, CTH, CER), zenith angles, surface temperatures (ST), RF values, and contrail sizes of all contrails detected over the two-week period.}
    \label{corr}
\end{figure}

For each contrail detected during the two-week period, we evaluated the linear relationship between its radiative properties and various factors, including contrail characteristics (such as size), surface temperatures, and lighting conditions (as shown in Figure \ref{corr}).
\\\\The key insights of the analysis include:\\ 

\begin{figure*}[!t]
    \centering
    \includegraphics[scale=0.4]{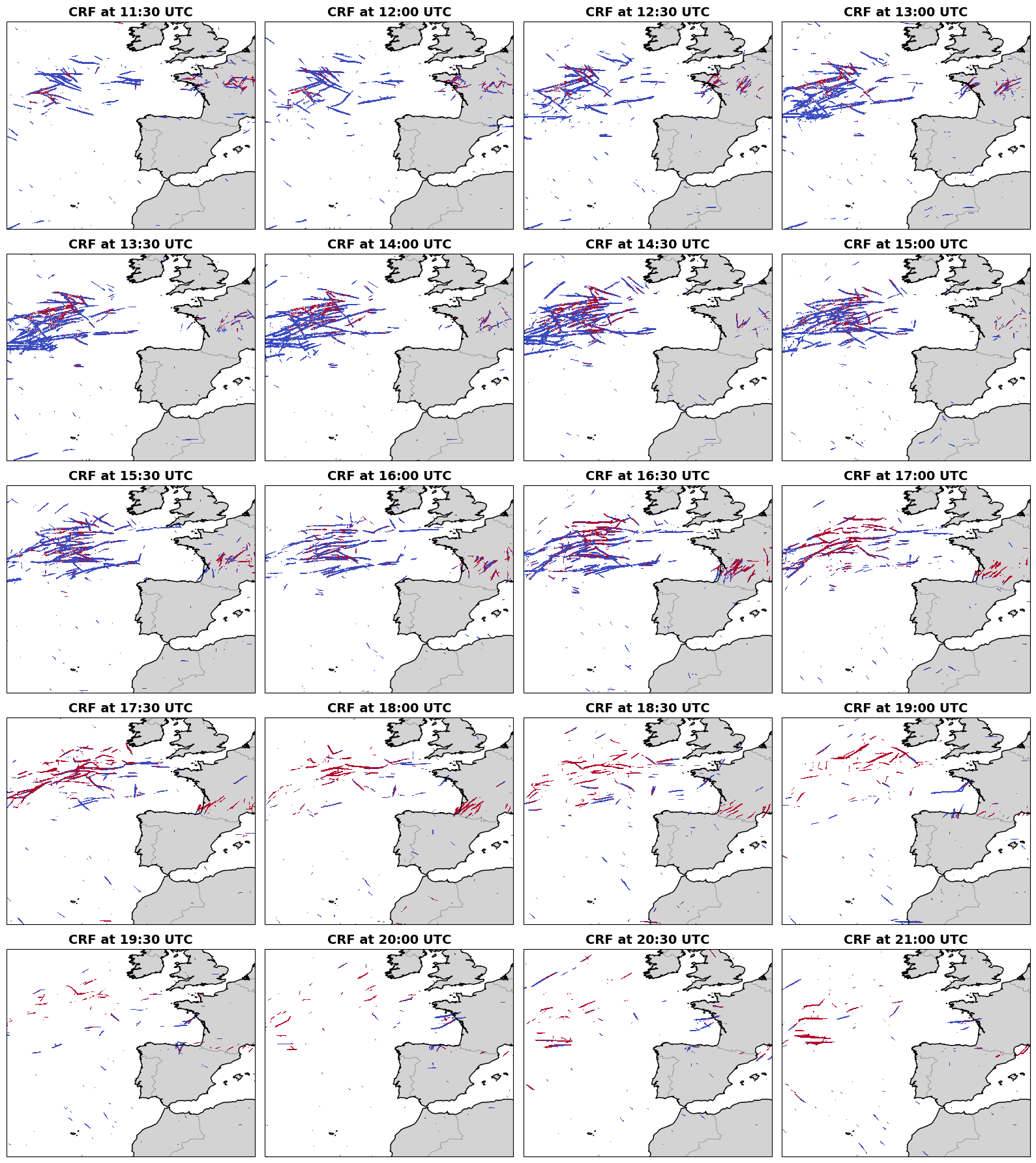}
    \caption{Contrail Outbreak over the Atlantic Ocean during a 9.5-hour period on January 30, 2023. The red colors represent positive Radiative Forcing (RF) values, indicating a contrail warming effect, while the blue colors represent negative RF values, indicating a contrail cooling effect. In this context, CRF stands for Contrail Radiative Forcing and UTC refers to Coordinated Universal Time.}
    \label{outbreak}
\end{figure*}

\begin{itemize}
\item The Net RF (W), which represents all the reflected and emitted radiation by the contrail, does not have a linear relationship with its size. Larger contrails exhibit both stronger warming and cooling (under shortwave radiation), balancing each other out. This strong dependency on the shortwave forcing suggests that nighttime contrails, without cooling, have the greatest warming impact, regardless of size. In other words, this means that \textit{a long, thick contrail formed during the day has a smaller warming impact compared to a small, thin contrail formed at night.} Figure \ref{outbreak} illustrates a large outbreak with a net cooling effect during the day, which, after sunset, transitions to a few small warming contrails.\\
\item The Average RF (W/m$_2$), which measures radiation at a single point of a contrail, is strongly correlated with the zenith angle, indicating that lighting conditions primarily determine whether a contrail warms the Earth. This is significant, as it suggests that \textit{most contrails appear to have a cooling effect during daytime}. Figure \ref{outbreak} illustrates the evolution of the contrail warming effect as daylight decreases. Average RF is also influenced, though to a lesser extent, by cloud altitude (CTH), with  \textit{higher-altitude contrails generally trapping more heat}.\\
\item It is important to note that our analysis focused solely on linear relationships. To gain a more comprehensive understanding of the factors affecting the warming and cooling impacts of individual contrails, additional effects must be considered. For example, the influence of underlying water clouds was not included in this analysis, as linear correlations with radiative forcing were found to be negligible. However, the interactions between contrails and underlying clouds, as well as between different contrail features, are known to involve more complex, non-linear processes. Previous studies have demonstrated that these contrail-cloud and contrail-contrail interactions have a significant impact on the total radiative forcing of a contrail \cite{sanz-morere_effect_2020} and should therefore be considered and included in our future analysis. 
\end{itemize}
\begin{figure*}[!t]
    \centering
    \includegraphics[scale=0.44]{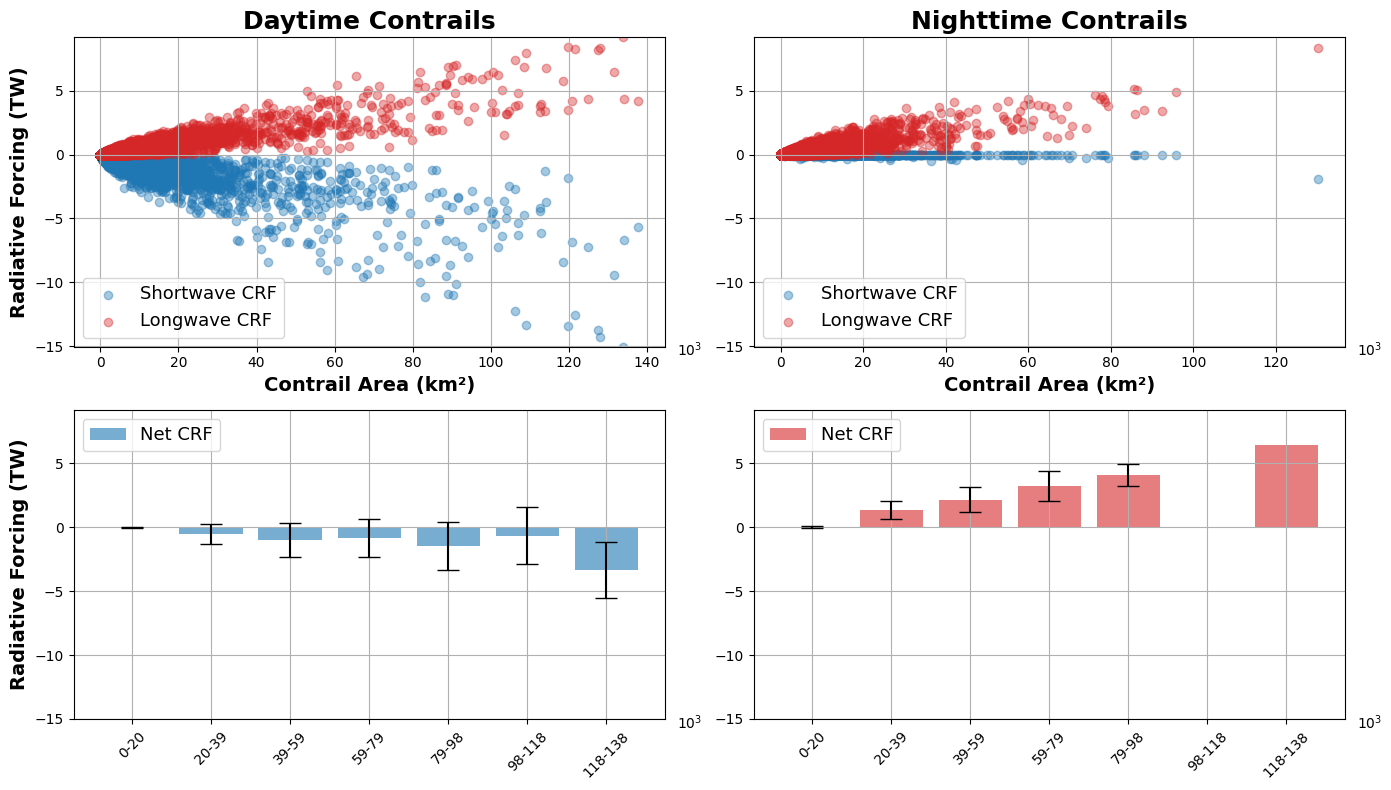}
    \caption{The top row displays the Radiative Forcing (RF) of contrail features by size, measured in terawatts, with separate panels for daytime (left) and nighttime (right). The bottom row shows the Short Wave (SW) and Long Wave (LW) RF components for daytime (left) and nighttime (right), which were combined to derive the RF values presented in the top row.}
    \label{rf_single_contrails}
\end{figure*}
\subsection{Day-Night Patterns}
Given that lighting conditions significantly influence a contrail's overall effect, we analyze each contrail's warming contribution by size, distinguishing between daytime and nighttime contrails. The top row bar plots in Figure \ref{rf_single_contrails} show that daytime contrails generally produce a cooling effect. Even the largest daytime contrails typically don't warm as much as an average-sized nighttime contrail. Given that most daytime contrails exhibit a cooling effect, a significant factor contributing to the overall warming effect throughout the day (Table \ref{rf_comparison}) is the higher proportion of nighttime contrails. In particular, only 38 $\%$ of all contrails detected over the 14-day period occurred during daylight, with the remaining 62$\%$ were observed at night.

\section{Conclusions}
This study evaluates the radiative impact of all contrails detected in SEVIRI data, using a neural network, over a two-week period spanning January 24th-30th in both 2023 and 2024. Detected daytime contrails generally produce a cooling effect, with minimum net CRF values reaching approximately -8 TW, while nighttime contrails contribute to warming, with CRF values up to 6 TW. Although most detected daytime contrails exhibit a cooling effect, the overall daily impact is warming due to the higher frequency of detected nighttime contrails, which constitute 62\% of the total detected. A comparison of the 2023 and 2024 data reveals a substantial increase in contrail coverage, with detections rising by 41.03\% and CRF values increasing by 128.7\% in 2024, indicating an intensified warming effect linked to the greater number of contrails. Analysis of individual contrail characteristics suggests that larger daytime contrails have a smaller warming impact than smaller nighttime contrails, reinforcing the overall warming trend. The study underscores the significant role of contrail timing in their net warming effect, highlighting the importance of considering both daytime and nighttime contrails in radiative forcing assessments. Understanding these temporal differences is essential for accurately evaluating the influence of contrails on climate change and for developing effective strategies to mitigate aviation-induced warming. In future work, a full year of data will be analyzed to capture seasonal variations and account for fluctuations in flight patterns, offering a more comprehensive assessment of contrail impacts throughout the year. Additionally, comprehensive validation of detected contrails or the introduction of uncertainty metrics will be undertaken to address potential errors and enhance the accuracy of findings.

\section*{Data Availability}
The LUTs, which are employed to compute the shortwave and longwave radiative forcing values for all clouds within the imagery, will soon be made available to the research community, accompanied by a forthcoming publication. Additionally, the models utilized for detecting contrail features withn MSG images will also be released to the community, with an accompanying publication currently in preparation. 

\section*{Contributions of the Authors}
IO is the principal author, she has developed the AI detection algorithm (supported by AJ and supervised by JGH and MS), has done the analysis of the data and has acted as the primary writer of the document. ED, PB, and NC developed the RF estimation algorithm, provided MSG data, and wrote the parts of the paper devoted to it. JGH, HB, JvG have participated in the data handling, processing, and post-processing. JGH also led the experimental plan. KS, EO and PL have participated in all the discussions related to data analysis and integration. MS is the coordinator of the project, he conceived the idea of the project, and the experimental setup (with the close collaboration of JGH and the rest of the members of the consortium).
\section*{Acknowledgment}

This research has been conducted within E-CONTRAIL project, an exploratory research project funded by the SESAR 3 Joint Undertaking and its members under grant agreement No 101114795. We would like to extend our sincere gratitude to all members of the E-CONTRAIL project for their invaluable contributions. Special thanks to Gracia Pérez, E-CONTRAIL Project Manager, and to Virginia Villaplana, E-CONTRAIL Communication manager, for her extensive assistance in all the E-CONTRAIL activities.
%

\bibliographystyle{IEEEtran}
\bibliography{references2}

\end{document}